\documentclass[aps,prb,twocolumn]{revtex4}
\usepackage{graphicx}
\usepackage{bm}
\usepackage{amssymb}
\usepackage{amsmath}
\usepackage{xcolor}
\usepackage{verbatim}
\newcommand{\da}{^{\dagger}}
\newcommand{\doa}{\downarrow}
\newcommand{\upa}{\uparrow}

\newcommand{\bvec}{\mathbf} 
\newcommand{\kl}{_{\mathbf{k}\lambda}}
\newcommand{\on}{^{(1)}}
\newcommand{\mon}{^{(-1)}}
\newcommand{\bveck}{\mathbf{k}}
\newcommand{\bveckp}{\mathbf{k'}}

\newcommand\be            {\begin{equation}}
\newcommand\ee            {\end{equation}}
\newcommand\ba            {\begin{aligned}}
\newcommand\ea            {\end{aligned}}
\newcommand\bvp{\mathbf{p}}
\newcommand\bvq{\mathbf{q}}
\usepackage{tikz}
\usetikzlibrary{positioning,arrows}
\usetikzlibrary{decorations.pathmorphing}
\usetikzlibrary{decorations.markings}
\tikzset{every picture/.style=thick}
 \tikzset{
f/.style={thick,draw=black, postaction={decorate},
    decoration={markings,mark=at position .52 with {\pgftransformscale{1.2}\arrow[black]{stealth}}}},
n/.style={decorate, draw=black,postaction={decorate},
    decoration={snake,amplitude=.3mm,segment length=1.8mm,post length=.0025mm}},
u/.style={draw=black},
arr/.style={thick,draw=black, postaction={decorate},
    decoration={markings,mark=at position 1 with {\pgftransformscale{1.1}\arrow[black]{stealth}}}},
 }
\pgfdeclarelayer{edgelayer}
\pgfdeclarelayer{nodelayer}
\pgfsetlayers{edgelayer,nodelayer}

\begin{document}

\title{Rashba versus Kohn-Luttinger: Evolution of $p$-wave superconductivity in magnetized two-dimensional Fermi gases subject to spin-orbit interactions}
\author{Ethan Lake}
\author{Caleb Webb}
\author{D. A. Pesin}
\author{O. A. Starykh}
\affiliation{Department of Physics and Astronomy, University of Utah, Salt Lake City, UT 84112, USA}

\date{\today}

\begin{abstract}
We study how the Rashba spin-orbit interaction influences unconventional superconductivity in a two dimensional electron gas partially spin-polarized by a magnetic field.
Somewhat surprisingly, we find that for all field orientations, only the larger Fermi surface is superconducting. When the magnetic field is oriented out-of-plane the
system realizes a topological $p+ip$ pairing state. When the field is rotated in-plane the order parameter develops nodes along the field direction and
finite center-of-mass-momentum pairing is realized. We demonstrate that the pairing symmetry of the system can be easily probed experimentally due to
the dependence of various thermodynamic quantities on the magnetic field geometry, and calculate the electronic specific heat as an example.
\end{abstract}

\maketitle

\section{Introduction}

The idea that a superconducting state can arise from direct repulsive interactions between electrons was first introduced in a celebrated paper by Kohn and Luttinger \cite{Kohn65} (see \cite{kagan2015,Maiti2013a} for recent reviews on this topic). Even though the bare interaction $U$ between electrons is repulsive, an effective attraction arises at $O(U^2)$ which forms a $p$-wave superconducting state in three dimensions. Unfortunately this type of pairing is much weaker in two dimensional systems, and in the weak-coupling regime only a fragile superconducting state at $O(U^3)$ exists\cite{Chubukov93}.

However, partially polarizing a two dimensional electron gas (2DEG) with an in-plane magnetic field can dramatically increase the strength of the effective pairing interaction at $O(U^2)$, as has been shown by a perturbative diagrammatic expansion \cite{Kagan1989} and more recently by an asymptotically exact renormalization group approach\cite{Raghu10,Raghu11}. It was found that a non-unitary $p+ip$ superconducting state forms, in which the larger of the 2DEG's two energy bands superconducts, while the smaller one does not.

This interesting result represents one of the conceptually simplest illustrations of the Kohn-Luttinger mechanism of superconductivity. The natural association
of the $S^z=+1$ order parameter with the majority (spin up) band ensures its `exotic' nature - the spatial part of the pair wave function must be an odd
function of the momentum (an odd angular harmonic). The fact that the highest critical temperature is achieved for the smallest possible $\ell=1$
harmonic \cite{Kagan1988} brings this idealized set-up very close to the forefront of modern research in topological states of matter \cite{Alicea2012}.

However, any physical discussion of two-dimensional superconductivity requires one to account for an omnipresent spin-orbit interaction
\cite{Gor'kov01,Barzykin02,Agterberg2007,Dimitrova2007,Michaeli2012,Loder13,Wang14}.
In this paper, we extend previous studies by investigating the effects of Rashba spin-orbit coupling (SOC) on unconventional triplet superconductivity,
mediated by repulsive electron interactions, in a partially spin-polarized 2DEG. We treat the SOC strength perturbatively with respect to the magnetic field strength,
while allowing the magnetic field to point at an arbitrary angle with respect to the plane of the electron motion.
As an example, such a scenario arises naturally at the boundary between LaAlO$_3$ and SrTiO$_3$\cite{Reyren07,Sing09}, where the spin-orbit interaction can be tuned by controlling an applied gate voltage\cite{Thiel06,Shalom10,Sachs10}.

The addition of SOC breaks spin conservation and induces coupling between spin-up and spin-down states.
It was speculated previously \cite{Raghu11} that SOC would therefore induce a superconducting state on the minority band due to the additional inter-band interactions it generates.
However, we find that {\it no} superconductivity is induced on the minority band to the leading order in the spin-orbit coupling strength, regardless of the mutual orientation of the magnetic field and
spin-orbital axes. A state where momentum-space Josephson coupling induces superconductivity on the minority band does exist, but it is energetically disfavored compared to a decoupled phase where only the majority band is superconducting.

We analyze the pairing symmetries realized on the majority band as a function of the magnetic field orientation. For as long as the magnetic field possesses a finite out-of-plane component the majority band is fully gapped with $p+ip$ symmetry, albeit with an angle-dependent, modulated gap. When the field is oriented strictly in-plane, the order parameter develops nodes along the field direction. Moreover, we find that quite generally the emergent superconducting state is of FFLO kind,
with a finite center-of-mass pair momentum \cite{FF64,LO65,Casalbuoni04}, which is determined by the vector product of the spin-orbital and magnetic fields.

Studying the details of the band structure and the direction of the nodes has been an area of intensive focus in recent experimental studies of unconventional superconductivity\cite{Matsuda06,Hashimoto09,Zeng10,Wu10}. We show that the dependence of the pairing symmetry on the magnetic field orientation allows the nodal symmetries of the system to be readily probed by measurements of various thermodynamic quantities. To illustrate this, we compute the electronic specific heat of the system as a function of the magnetic field orientation.

The structure of this paper is as follows. In section~\ref{sec:hamiltonian} we introduce the Hamiltonian and perform a unitary transformation which brings it into a form diagonal in the band indices. Self-consistent equations of the theory are derived in section~\ref{sec:mf}.
In section~\ref{sec:OPs} we detail the resulting order parameter symmetries for different magnetic field orientations. Calculation of the electronic specific
heat in section~\ref{sec:specific_heat} is followed by the discussion of our results in section~\ref{sec:dis}.
A complementary discussion of finite center-of-mass momentum pairing is presented in Appendix A. 

\section{Model and Hamiltonian} \label{sec:hamiltonian}

\subsection{Hamiltonian}

\begin{figure}
\includegraphics{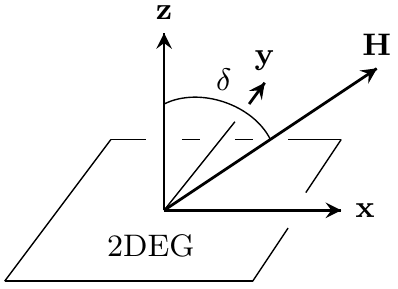}
\caption{\label{fig:setup} Geometry of the system. We align the spin quantization axis along magnetic field $\bvec{H}$.}
\end{figure}

The geometry of the system we consider is shown in Figure~\ref{fig:setup}. We let the 2DEG lie in the $xy$ plane, with the magnetic field $\bvec{H}$ inclined
by a polar angle $\delta$ relative to the $\bvec{\hat z}$ axis with azimuthal angle $\phi = 0$.
The non-interacting part of the Hamiltonian $H_0$ describes electrons with parabolic dispersion $\bvec{k}^2/(2m)$ subject to the external
Zeeman field $-g\mu_B\bvec{H}\cdot\pmb{\sigma}/2$, where $g$ is the $g$-factor and $\mu_B$ is Bohr magneton, and the spin-orbit interaction of Rashba type
$\alpha_R \bvec{k}\times\pmb{\sigma}\cdot \hat{z} = \alpha_R (k_x \sigma^y - k_y \sigma^x)$. We set $\hbar = c = 1$ throughout the paper.

In order to conveniently treat SOC in perturbation theory, we choose to align the spin quantization axis parallel to the magnetic field.
This is done with the help of a unitary rotation about the $\hat{y}$ axis, ${\cal R}_y = \exp[-i\delta \sigma^y/2]$, which transforms $H_0$ into
\begin{equation}
H_0  = \sum_{\bveck\sigma\sigma'} E_{\sigma\sigma'}(\bvec{k}) c\da_{\bveck\sigma}c_{\bveck\sigma'},
\end{equation}
where the matrix $E(\bveck)$ is given by
\begin{equation} \label{eq:E}
E(\bveck) = \frac{k^2}{2m}\sigma^0 - h\sigma^z + \alpha_R(k_x\widetilde\sigma^y - k_y\widetilde\sigma^x)
\end{equation}
and $\boldsymbol{\widetilde\sigma} = \left(\sigma^z\sin\delta + \sigma^x\cos\delta\right) \bvec{\hat x} + \sigma^y\bvec{\hat y} + \left(\sigma^z\cos\delta - \sigma^x \sin\delta\right)\bvec{\hat z}$ represents the rotated spin $\pmb{\sigma}$.
The Zeeman coupling strength is given by $h = eg|\bvec{H}|/2m$. We neglect the effect of the external field on the orbital motion of the electrons, an approximation which is justified if the $g$-factor is sufficiently large or if the orbital coupling term is absent. This is the case in systems of cold neutral atoms, which have been the subject of several recent experimental studies of SOC\cite{Lin11,Cheuk12,Qu13a}.

The electrons repel each other via the short-ranged (contact) interaction $H_I = \int d^2 {\bf r} ~U \hat{n}_\uparrow({\bf r}) \hat{n}_\downarrow({\bf r})$
written in terms of electron spin densities $ \hat{n}_\sigma({\bf r})$.
Therefore the full Hamiltonian for our system in momentum space is
\begin{equation}
\begin{aligned} \label{eq:H}
H & = H_0 + H_I, \\
H_0 & = \sum_{\bveck\sigma\sigma'} E_{\sigma\sigma'}(\bvec{k}) c\da_{\bveck\sigma}c_{\bveck\sigma'} \\
H_I & = \frac{U}{V}\sideset{}{'}\sum_{\bveck_1\bveck_2\bveck_3\bveck_4} c\da_{{\bveck}_1\upa}c\da_{{\bveck}_2\doa}c_{{\bveck}_3\doa}c_{\bveck_4\upa}
\end{aligned}
\end{equation}
where the primed sum is subject to the momentum conservation $\bvec{k}_1 + \bvec{k}_2 = \bveck_3 + \bveck_4$.

\subsection{Schrieffer-Wolff transformation}

In this Section we describe the single particle spectrum of Hamiltonian (\ref{eq:H}), and construct a canonical (Schrieffer-Wolff) transformation,
which brings the interaction Hamiltonian into a form convenient for mean-field analysis.

The Hamiltonian (\ref{eq:H}), for general values of $\alpha_R$, $h$ and $U$, is extremely complicated, and is amenable only to numerical treatments.
In what follows we make several physically-motivated simplifying assumptions, which restrict the generality of the obtained results,
but make the problem analytically solvable. Throughout, we assume the spin-orbit interaction to be weak compared to the Zeeman coupling, allowing us to
treat the ratio $\alpha_R k_f/h$ perturbatively, where $k_f$ is the Fermi momentum of either Fermi surface. The particle-particle interaction, $U$, is assumed to be weak, $mU\ll1$.
As far as quantities of higher order of smallness are concerned, we will keep terms of order $O(mU\alpha_R^2k_f^2/h^2)$ while discarding those of order $O(m^2U^2\alpha_Rk_f/h)$, which is
permissible for not too small Rashba SOC, $\alpha_R k_f\gg mU h$. The utility of these approximations will become clear in what follows.

We begin with the single-particle part of the Hamiltonian, $H_0$. The problem of finding the spectrum and eigenstates of $H_0$ can be easily solved exactly; however, for our purposes we specialize to the case of weak SOC from the outset.

To diagonalize $H_0$, we perform a unitary transformation from the operators $c_{\bf{k}\sigma}$ to band operators $a\kl$, where the index $\lambda\in \{1,2\}$ labels the two bands, with $1$ denoting the larger (majority) band and $2$ the smaller (minority) ones. To the required order in $\alpha_R$, the transformation to band operators is given by
\begin{equation} \label{eq:band_tform}
\begin{aligned}
& c_{\bveck\upa} = \left[1 - \frac{\alpha_R^2k^2}{8h^2}(\cos^2\phi_{\bveck} + \sin^2\phi_{\bveck} \cos^2\delta)\right]a_{\bveck1} \\ & \qquad\qquad - \frac{\alpha_R k}{2h}(i\cos\phi_{\bveck} + \sin\phi_{\bveck}\cos\delta) a_{\bveck2},\\
& c_{\bveck\doa} = -\frac{\alpha_R k}{2h}(i\cos\phi_{\bveck} - \sin\phi_{\bveck}\cos\delta) a_{\bveck1} \\ & \qquad\qquad + \left[1 - \frac{\alpha_R^2k^2}{8h^2}(\cos^2\phi_{\bveck} + \sin^2\phi_{\bveck}  \cos^2\delta)\right]a_{\bveck2},
\end{aligned}
\end{equation}
where $\phi_{\bveck}$ is the azimuthal angle of ${\bveck}$.

The exact band dispersion is calculated from Eq.~(\ref{eq:E}) as
\begin{equation} \label{eq:disp}
\varepsilon_{\lambda}(\bveck) = \frac{k^2}{2m} - \zeta_{\lambda} \sqrt{\alpha_R^2 k^2 + h^2 + 2\alpha_R h k \sin\phi_{\bveck} \sin\delta}
\end{equation}
with $\zeta_{\lambda}$ denoting the helicity of each band defined with respect to the $\hat z$ axis. We use $\zeta_1 = 1$ ($\zeta_2 = -1$) for the majority (minority) band.

It is worth noting that to the leading order in the expansion parameter $r=\alpha_R k_f/h \ll 1$ the obtained band dispersion can be approximated as
\begin{equation} \label{eq:disp2}
\varepsilon_{\lambda}(\bveck) \approx \frac{k_x^2 + (k_y - \zeta_\lambda Q)^2 - Q^2}{2m} - \zeta_{\lambda} h,
\end{equation}
where $Q = m \alpha_R \sin\delta$ describes the center-of-mass momentum shift of the two bands. Observe that to this order the dispersion
retains circular shape, and that the bands shift in the {\em opposite} directions, see Figure~\ref{fig:FS}.
Finally, note that vector ${\bf Q} = Q \hat{y} \propto \hat{z} \times {\bf H}/|{\bf H}|$ is orthogonal to the ${\hat z}-{\hat x}$ plane.

\begin{figure}
\includegraphics{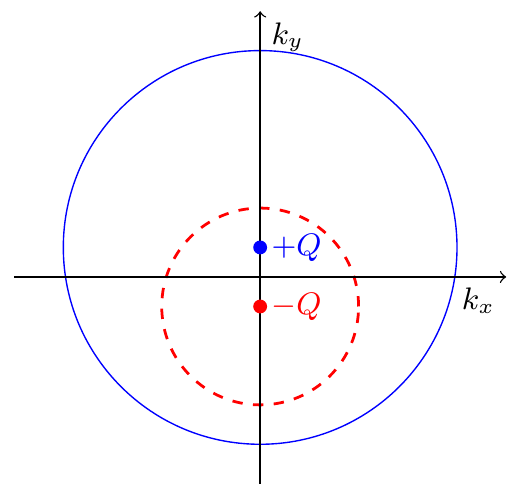}
\caption{Schematics of the Fermi surface geometry when the magnetic field has a finite in-plane component along $\hat{x}$.
Fermi surface of the majority (minority) bands is shown in solid/blue (dashed/red). Filled points denote the location of their centers.}
\end{figure}

Next we turn to the interaction Hamiltonian $H_I$. Our approach is motivated by the observation that in the absence of SOC, short-range repulsion is of purely interband character: particles with the same spin orientation feel no local interaction due to the Pauli exclusion principle. One may then utilize the Schrieffer-Wolff transformation\cite{LossSWT} to get rid of this interband interaction in favor of an effective intraband one.  We will accomplish a similar program in the presence of SOC. It is known that without SOC, the effective intraband interaction leads to $p$-wave superconductivity on the majority Fermi surface\cite{Chubukov93,Raghu11}. We show below that in the presence of the Rashba SOC the intraband attraction is supplemented by two more terms:  an intraband repulsion as well as an interband Josephson coupling.

As a first step, we recast the interaction Hamiltonian $H_I$ in terms of the band operators $a\kl$. As was mentioned before, we
keep the leading interactions that go as $O(U^2)$, and $O(U\alpha_R^2)$, but drop those that go as $O(U^2\alpha_R)$.

With this prescription we find that $H = H_0 + H_1 + H_2$,
\begin{equation}
\begin{aligned}
H_0 & = \sum_{\lambda} \sum_{\bveck} \varepsilon\kl a\da_{\bveck\lambda}a\kl,
\end{aligned}
\end{equation}
and where, schematically, $H_1$ contains all terms of the form $a\da_{\lambda}a\da_{\lambda}a_{\mu}a_{\mu}$, and
$H_2$ contains all terms of the form $a\da_{\lambda}a\da_{-\lambda} a_{-\lambda}a_{\lambda}\propto n_\lambda n_{-\lambda}$
(with $-\lambda$ denoting the opposite to $\lambda$ band index) as well as those
with three operators of the same band index.

Since we are only concerned with the interaction in the Cooper channel, the interaction Hamiltonian $H_1$ describes intraband interaction in this channel -- repulsion
(for $\mu=\lambda$), as well as interband Josephson coupling (for $\mu=-\lambda$). Both types of terms are of $O{(U\alpha_R^2)}$ order.
The interaction terms in $H_2$ are off-diagonal in the band index, but give effective intraband attraction at $O(U^2)$ order\cite{Chubukov93,Raghu11}.
Therefore, our strategy is to perform a unitary transformation on the Hamiltonian in order to eliminate $H_2$ from the resulting expression, in favor of an effective intraband interaction. The transformed Hamiltonian is
\begin{equation}
\widetilde H = e^{-S}He^S
\end{equation}
where $S = -S\da$ is anti-Hermitian, leaving the eigenvalues of the Hamiltonian unchanged. We choose the ansatz $[S,H_0] = H_2$, and to the level of approximation stated earlier we obtain
\begin{equation}\label{eq:Hrotated}
\begin{aligned}
\widetilde H = H_0 + H_1 + [H_1,S] + \frac{1}{2}[H_2,S].
\end{aligned}
\end{equation}

It is straightforward to show that $S$ is given by the sum of each term in $H_2$ divided by the energy exchange mediated by each term. For example, for the term in $H_2$ of the form
\begin{equation}
\frac{U}{V}\sum_{\bveck_1\bveck_2\bveck_3} a\da_{\bveck_1,\lambda}a\da_{\bveck_2,-\lambda}a_{\bveck_3,-\lambda}a_{\bveck_4,\lambda}
\end{equation}
$S$ contains the corresponding term
\begin{equation}
\label{eq:S}
\frac{U}{V}\sum_{\bveck_1\bveck_2\bveck_3} \frac{a\da_{\bveck_1,\lambda}a\da_{\bveck_2,-\lambda}a_{\bveck_3,-\lambda}a_{\bveck_4,\lambda}}{\varepsilon_{\bveck_4,\lambda} + \varepsilon_{\bveck_3,-\lambda} - \varepsilon_{\bveck_2,-\lambda} - \varepsilon_{\bveck_1,\lambda}}.
\end{equation}

\begin{figure}
\includegraphics{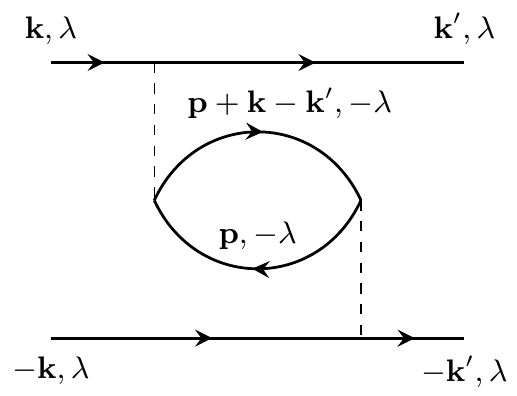}
\caption{\label{fig:KL_bubble} The diagram responsible for creating a pairing instability at second order in the particle-particle interaction strength. Two electrons on band $\lambda$ with momenta $\pm \bveck$ scatter to momenta $\pm \bvec{k'}$, creating a particle-hole pair on the opposite band $-\lambda$. }
\end{figure}

Using the fact that $H_1 \propto O{(U\alpha_R^2)}$, as explained above, we observe that $[H_1,S] \propto  O{(U^2\alpha_R^2)}$ and as such
represents a higher order correction to our leading order approximation. Hence we are allowed to drop it.
Commuting \eqref{eq:S} with $H_2$ in \eqref{eq:Hrotated} we generate three-particle (six $a$) terms where four $a$ operators
belong to the band $\lambda$, while the other two are from the opposite, $-\lambda$, band. Therefore the effective intraband interaction
in the $\lambda$-band is obtained by {\em projecting} all $a_{-\lambda}$ operators to the non-interacting $-\lambda$ band.
Such a projection turns out to be equivalent to the replacement of number conserving operator products by the appropriate
Fermi distributions: $a^\dagger_{{\bf k},-\lambda} a_{{\bf k}',-\lambda} \to \delta_{{\bf k},{\bf k}'} f(\epsilon_{{\bf k},-\lambda})$,
where $f$ denotes the Fermi distribution.
Specializing next to the BCS (pairing) channel in the $\lambda$-band,
we find that the last term in \eqref{eq:Hrotated} produces an effective interaction between electron pairs on the Fermi surface of the $\lambda$ band,
\begin{equation}
\label{eq:H-KL}
H_{\rm KL} = \frac{U^2}{2V} \sum_\lambda \sum_{\bveck,\bveck'} \chi_{-\lambda}({\bveck}-{\bveck'}) a^\dagger_{-{\bf k}',\lambda} a^\dagger_{{\bf k}',\lambda}
a_{{\bf k},\lambda} a_{-\bveck,\lambda}.
\end{equation}
$H_{\rm KL}$ captures the Kohn-Luttinger physics, and describes the $\lambda$-band pairing interaction generated by the
particle-hole fluctuations in the opposite, $-\lambda$, band. This result is fully equivalent to previous diagrammatic calculations \cite{Chubukov93,Raghu11}
and is graphically described by the diagram in Figure \ref{fig:KL_bubble}.
The functions $\chi_{\lambda}(\bveck-\bveck')$ are the particle-hole susceptibilities for each band, defined by
\begin{equation}\begin{aligned}\label{eq:chi}
\chi_{\lambda}(\bvec{q}) = \frac{1}{V} \sum_{\bvec{p}} \frac{f(\varepsilon_{\bvec{p}+\bvec{q},\lambda}) - f(\varepsilon_{\bvec{p},\lambda})}{ \varepsilon_{\bvec{p}+\bvec{q},\lambda} - \varepsilon_{\bvec{p},\lambda}}.
\end{aligned} \end{equation}

Adding $H_1$ contributions to \eqref{eq:H-KL}  we finally obtain for the transformed Hamiltonian \eqref{eq:Hrotated}
\begin{equation}
\begin{aligned} \label{eq:tformedH}
\widetilde H = H_0 +\sum_{\lambda\mu} \sum_{\bveck,\bveck'} g_{\lambda\mu}(\bveck,\bveck') a^\dagger_{-{\bf k}',\lambda} a^\dagger_{{\bf k}',\lambda}
a_{{\bf k},\mu} a_{-\bveck,\mu},
\end{aligned}
\end{equation}
where the full interaction matrix, $g_{\lambda\mu}(\bveck,\bveck')$, is given by
\begin{equation} \label{eq:intmat}
\begin{aligned}
& g_{\lambda\mu}(\bveck,\bveck') = \frac{U\alpha_R^2k_{{\rm f},\lambda}k_{{\rm f},\mu}\zeta_{\lambda}\zeta_{\mu}}{4Vh^2}(\cos\phi_{\bveck} - i\zeta_{\lambda}\cos\delta\sin\phi_{\bveck}) \\ & \qquad \times (\cos\phi_{\bveck'} + i\zeta_{\mu}\cos\delta\sin\phi_{\bveck'}) + \frac{U^2 \chi_{-\lambda}({\bveck}-{\bveck'})}{2V}\delta_{\lambda\mu} \\
\end{aligned}
\end{equation}
where $k_{{\rm f},\lambda}$ is the Fermi momentum on each band, $\phi_{\bveck}$ is the angle momentum $\bveck$ makes with ${\hat x}$-axis,
and, as before, $\zeta_{\lambda} = 1$ ($-1$) for $\lambda = 1$ ($\lambda = 2$)
is the helicity of each band.

The first term in $g_{\lambda\mu}(\bveck,\bveck')$ represents the interactions generated by SOC and comes from $H_1$ contributions.
The factor of $\zeta_{\lambda}\zeta_{\mu}$ tells us that the intra-band interactions generated by SOC are repulsive,
while the inter-band Josephson terms are attractive.

In writing  the Hamiltonian \eqref{eq:tformedH} we neglected the shifts of the Fermi surfaces away
from their zero-momentum centers, $\zeta_\lambda Q$.
Keeping these shifts leads to small corrections beyond the assumed approximation described in the beginning of this subsection.
A more complete form of the Hamiltonian, which takes into account these shifts, is derived in Appendix A (see \eqref{eq:app1}),
where details of pairing with finite momentum are discussed.

To compute the susceptibility, we set $\alpha_R = 0$ in (\ref{eq:chi}), which is permissible since $\chi_\lambda$ is multiplied by $U^2$ in the interaction matrix,
and therefore the SOC-induced corrections are $O(U^2\alpha_R)$ and can be dropped. This is one of the major simplifications that come from our assumptions.
At the same time, it is worth noting that the rather special (circular) form of the dispersion \eqref{eq:disp2} guarantees that
the SOC-induced corrections to $\chi_{\lambda}(\bvec{q})$ can only appear in $O(\alpha_R^2)$ or higher order because the transferred momentum
$\bvec{q} = \bvec{k} - \bvec{k'}$ does not depend on the overall momentum shift $\zeta_\lambda Q$ as long as the momenta $\bvec{k},\bvec{k'}$
belong to the Fermi surface of the same band.

Integrating \eqref{eq:chi} gives an explicit expression for the charge susceptibility\cite{Raghu11}:
\begin{equation}\label{eq:chifull}
\chi_{\lambda}(\bvec{q}) = -\rho_{\lambda}\left(1 - \frac{\textrm{Re}\sqrt{q^2 - (2k_{{\rm f},\lambda})^2}}{{q}}\right)
\end{equation}
where $\rho_{\lambda}$ is the density of states of the band $\lambda$. Again, the exact expression for $\rho_{\lambda}$ is band and momentum dependent, but
the corrections to $\rho_{\lambda} = m/2\pi$ are $O(\alpha_R)$ and hence need to be dropped when they meet the factor of $U^2$ present in the interaction matrix.

The fact that $\chi_{\lambda}(\bvec{q})$ reduces to a constant for $q \leq 2 k_{{\rm f},\lambda}$ is of major importance for our problem.
It implies that electrons on the minority ($\lambda =2$) Fermi surface do not experience any attractive potential from the particle-hole
fluctuations produced by the majority ($\lambda =1$) electrons. Indeed, the triplet nature of the electron pair on the majority/minority
Fermi surface (the total spin of the pair is $S^z = +1/-1$, correspondingly) requires the coordinate part of the pair wavefunction to be
an {\em odd} angular harmonic which, at the very minimum, requires that the interaction which mediates the attraction is momentum-dependent.
However, the maximum momentum transfer that minority electrons may experience
is $2 k_{{\rm f},2}$ which, by virtue of the finite magnetization $M \propto k_{{\rm f},1} - k_{{\rm f},2} > 0$,
is always less than the `critical' $2 k_{{\rm f},1}$ value needed in order for $\chi_{1}(\bvec{q})$ to acquire momentum dependence.
The same argument, when applied to the electron pairs on the majority Fermi surface, shows that $\chi_{2}(\bvec{q})$ is certainly
momentum dependent for $2 k_{{\rm f},2} < q < 2 k_{{\rm f},1}$. Therefore, triplet pairing is possible on the majority surface.

Furthermore, calculations \cite{Kagan1989,Raghu11} show that the pairing is strongest for the $p$-wave, $l = \pm 1$,
harmonic $\chi_2^{(\ell = \pm 1)}$ of $\chi_{2}$. One finds $\chi_2\on = \chi_2\mon = -\rho_2\eta(1-\eta)$,
where $\eta = k_{{\rm f},2} / k_{{\rm f},1} < 1$ is the ratio of the Fermi momenta.
We note that $\chi_2\on$ is most negative, {\it i.e.} most attractive, for $\eta = 1/2$.\cite{Kagan1989,Raghu11}

\section{Mean Field Theory} \label{sec:mf}

The mean-field approximation is formulated by introducing the order parameters for each band as
\begin{equation}\begin{aligned} \label{eq:opdef}
\Delta_{\lambda}(\bveck) = \sum_{\mu} \sum_{\bveck'} g_{\lambda\mu}(\bveck,\bveck')\langle a_{-\bveck'\mu}a_{\bveck'\mu}\rangle.
\end{aligned} \end{equation}

Using the Nambu spinor notation
\begin{equation}
\Psi\kl = (a\kl, ~a\da_{-\bveck\lambda})^T,
\end{equation}
the mean-field approximation for the Hamiltonian \eqref{eq:tformedH} reads as
\begin{equation}\begin{aligned}
H = \frac{1}{2}&\sum_{\lambda} \sum_{\bveck} \left(\Psi\da\kl \mathcal{H}_{\lambda}(\bveck) \Psi\kl + \xi\kl\right) \\ & - \sum_{\lambda\mu}\sum_{\bveck\bveck'}\Delta\da_{\lambda}(\bveck) g^{-1}_{\lambda\mu}(\bveck,\bveck') \Delta_{\mu}(\bveck')
\end{aligned} \end{equation}
where $\xi\kl = \varepsilon\kl - \mu$ and the matrix $\mathcal{H}_{\lambda}(\bveck)$ is represented by
\begin{equation}
\mathcal{H}_{\lambda}(\bveck) = \xi\kl\sigma^z + \Delta_{\lambda}(\bveck)(\sigma^x + i\sigma^y) + \Delta\da_{\lambda}(\bveck)(\sigma^x - i\sigma^y).
\end{equation}

The diagonalization of $\mathcal{H}_{\lambda}(\bveck)$ is accomplished by a straightforward Bogolyubov transformation, which at zero temperature gives a thermodynamic potential of
\begin{equation}\begin{aligned} \label{eq:omega}
\Omega = \frac{1}{2}\sum_{\lambda} &\sum_{\bveck} \left( \xi\kl - E\kl\right) \\ & - \sum_{\lambda\mu}\sum_{\bveck\bveck'}\Delta\da_{\lambda}(\bveck) g^{-1}_{\lambda\mu}(\bveck,\bveck') \Delta_{\mu}(\bveck'),
\end{aligned} \end{equation}
where the quasiparticle dispersion is
\begin{equation}\begin{aligned}
E\kl = \sqrt{\xi\kl^2 + 4|\Delta_{\lambda}(\bveck)|^2}.
\end{aligned} \end{equation}

We then derive the self-consistent equations for the theory by minimizing $\Omega$ with respect to each order parameter, obtaining
\begin{equation} \label{eq:sce} \Delta_{\lambda}(\bveck) = -\sum_{\mu} \sum_{\bveck'} g_{\lambda\mu}(\bveck,\bveck') \frac{\Delta_{\mu}(\bveck')}{E_{\bveck'\mu}}.\end{equation}

The condensation energy of the system
\begin{equation}
\begin{aligned} \label{eq:Ec}
E_c = & \frac{1}{2} \sum_{\lambda} \sum_{\bveck} (|\xi\kl| - E\kl) \\ & -\sum_{\lambda\mu} \sum_{\bveck\bveck'} \Delta\da_{\lambda}(\bveck)g^{-1}_{\lambda\mu}(\bveck,\bveck')\Delta_{\mu}(\bveck')
\end{aligned}
\end{equation}
can be significantly simplified further by using the explicit form of \eqref{eq:sce}:
\be
E_c = - \sum_{\lambda=1,2} \rho_\lambda |\Delta_{\lambda}(\bveck_{{\rm f},\lambda})|^2.
\ee

\subsection{Solutions and their symmetries}\label{sec:OPs}

Making use of (\ref{eq:intmat}) and (\ref{eq:sce}), the self-consistent equations read as
\begin{equation} \label{eq:SCsystem}
\begin{aligned}
& \Delta_{\lambda}(\bveck) = -\sum_{\mu,\bveckp} \frac{\Delta_{\mu}(\bveck')}{\sqrt{\xi_{\bveck'\mu}^2 + 4|\Delta_{\mu}(\bveck')|^2}}\bigg\{
\frac{\delta_{\lambda \mu} \delta_{\lambda 1}}{2}U^2\chi_{2}(\bveck-\bveck') \\
&+ \frac{U\alpha_R^2k_{{\rm f},\lambda}k_{{\rm f},\mu}\zeta_{\lambda}\zeta_{\mu}}{4h^2}(\cos\phi_{\bveck} - i\zeta_{\lambda}\cos\delta\sin\phi_{\bveck})\times\\
&\times (\cos\phi_{\bveck'} + i\zeta_{\mu}\cos\delta\sin\phi_{\bveck'}) \bigg\}
\end{aligned}
\end{equation}
Note that, as discussed in the end of section~\ref{sec:hamiltonian},
the Kohn-Luttinger mechanism provides only for the pairing in the majority ($\lambda=1$) band.
The $\chi_2(\bveck-\bveck')$ term is handled by restricting both momenta to the $\lambda=1$ Fermi surface, which is allowed
due to the presence of $\xi_{\bveck'}$ is the denominator of the right-hand side, and expanding
$\chi_2(\bveck_{{\rm f},1}-\bveck'_{{\rm f},1}) = \sum_{\ell\in {\rm odd}}\chi_2^{(\ell)} \cos[\ell(\phi_{\bveck}-\phi_{\bveck'})]$
in relative azimuthal angle. The strongest pairing is in the $\ell =1$ channel to which we restrict ourselves
in the following. (The coupling between different $\ell$ channels occurs only in higher orders of $\alpha_R$ expansion.)

We find that \eqref{eq:SCsystem} admits two distinct kinds of solutions which we call {\em coupled} and {\em decoupled}.

The coupled solution involves both bands and is characterized by the following angular structure
\begin{equation}\label{eq:coupled}
\begin{aligned}
& \Delta_{\lambda}(\bveck) = \Delta_{\lambda}(\cos\phi_{\bveck} - i\zeta_{\lambda}\cos\delta\sin\phi_{\bveck}),
\end{aligned}
\end{equation}
where the $\Delta_{\lambda}\neq 0$ are constants that must be solved for self-consistently. In this solution both bands are gapped, with momentum-space
Josephson coupling ($\propto U\alpha_R^2$) inducing a superconducting gap on the minority band. In addition, however, this solution is also characterized by the presence of a
repulsive intra-band interaction on each band ($\propto U\alpha_R^2(\cos^2 \phi' + \cos^2\delta \sin^2\phi) > 0$) that diminishes the attraction due to $\chi_2^{(1)}$ and the Josephson coupling.

Since $\Delta_{\lambda}$ is exponentially sensitive to the magnitude of the overall attractive interaction,
we are prompted to look for a pairing symmetry that minimizes their influence in the self-consistent equations.
Observe that the repulsive intra-band interactions on the majority band will vanish if $\Delta_1(\bveck)$ satisfies
\begin{equation} \label{eq:decoupled_condition}
\sum_{\bveck'} \frac{(\cos\phi_{\bveck'} + i\cos\delta\sin\phi_{\bveck'})\Delta_1(\bveck')}{\sqrt{\xi_{\bveck' 1}^2+4|\Delta_1(\bveck')|^2}} = 0.
\end{equation}
It is not difficult to see that (\ref{eq:decoupled_condition}) is satisfied if
\begin{equation} \label{eq:decoupledOP}
\Delta_1(\bveck) = \Delta_1(\cos\delta\cos\phi_{\bveck} + i\sin\phi_{\bveck}).
\end{equation}

In addition, if $\Delta_1(\bveck)$ satisfies (\ref{eq:decoupledOP}), then the inter-band Josephson coupling term in the minority-band self-consistent equation
vanishes as well. Since the only other term in the minority band self-consistent equation is repulsive, the energetically favorable solution is to have
$\Delta_2(\bveck) = 0$ everywhere. This {\em decoupled} solution thus realizes a phase where the minority band is not superconducting,
with the amplitude of the order parameter $\Delta_1$ on the majority band determined by solving
\begin{equation} \label{eq:decoupled_sce}
\Delta_1(\bveck) = - \sum_{\bveckp} \frac{U^2 \chi_{2}^{(1)}\Delta_1(\bveckp) \cos(\phi_{\bveck} - \phi_{\bveckp})}{2\sqrt{\xi_{\bveckp1}^2 + 4|\Delta_1(\bveckp)|^2}} .
\end{equation}
By combining \eqref{eq:decoupledOP} and \eqref{eq:decoupled_sce} we find
\begin{eqnarray}
&&\Delta_1(\bveck) =\Delta_1\big(\cos\delta \cos\phi_{\bveck} + i \sin\phi_{\bveck}\big),\nonumber\\
&&\Delta_1 =\frac{ 2\omega_c e^{\frac{1-\cos\delta}{2(1+\cos\delta)}}}{1+\cos\delta} \exp[-2/(U^2 \rho_1 |\chi_2^{(1)}|)],
\label{eq:decoupled-sol}
\end{eqnarray}
where $\omega_c$ is the standard upper-limit cutoff in the $\xi$ integration.

To determine which of the two found solutions is realized, we compare their condensation energies using Eq.~\eqref{eq:Ec}.
We find that for all values of the spin-orbit strength and all magnetic field orientations, the {\it decoupled}
solution has the lower condensation energy, and hence is physically realized. The basic reason for that is the aforementioned exponential sensitivity
of the energy gap to the magnitude of the attractive potential. This turns out to be a much stronger effect than the power-law gain ($\propto U\alpha_R^2$)
of the condensation energy due to the induced superconductivity in the minority band.

It is interesting to note that the self-consistent solution for the majority-band order parameter \eqref{eq:decoupled-sol} in the decoupled phase has no explicit
dependence on the spin-orbit interaction. This means that for small $\alpha_R$ the presence of a spin-orbit interaction affects the type of pairing symmetry
that is realized in the system, while the actual magnitude of the spin-orbit coupling does not influence the strength of the superconducting phase.
At the same time, $\Delta_1$ does depend on the magnetic field orientation via the $\cos\delta$ dependence in \eqref{eq:decoupled-sol}, and is
larger when the magnetic field is in-plane (at $\delta=\pi/2$).

A less-obvious feature of the found decoupled solution is that it is actually of the FFLO kind. Thanks to Eq.\eqref{eq:disp2} each member of the superconducting pair
carries finite momentum $Q \hat{y}$, resulting in the ${\bf Q}_{\rm pair}=2Q \hat{y}$ momentum of the pair. The fact that $\chi_2(\bveck)$ is not sensitive to $Q$,
as previously mentioned in the discussion above \eqref{eq:chifull}, makes the issue of the center-of-mass momentum `hidden' in the described analysis.
A two-particle Schrodinger equation re-formulation of the problem, outlined in the Appendix, makes this important point much clearer.
Our finding of finite ${\bf Q}_{\rm pair}$ is quite similar to that in Refs.\cite{Agterberg2007,Dimitrova2007} which considered the opposite from ours limit of strong
SOC, $\alpha_R k_{\rm f}/h \gg 1$.
Note that the fact that the two bands shift in {\em opposite} directions makes the coupled solution even more unfavorable -- under this condition the Josephson
coupling term does not conserve momentum, which has the effect of further suppressing it.

It is instructive to critically compare the symmetries of the decoupled and coupled phases for the limiting cases of
completely out-of-plane ($\delta = 0$) or completely in-plane ($\delta = \pi/2$) magnetic fields.

For $\delta = 0$ the coupled solution \eqref{eq:coupled} has $\Delta_1(\bveck) \propto e^{-i\phi_{\bveck}},~\Delta_2(\bveck) \propto e^{i\phi_{\bveck}}$
while the decoupled one \eqref{eq:decoupled-sol} has $\Delta_1(\bveck) \propto e^{i\phi_{\bveck}},~\Delta_2(\bveck) = 0$.
Thus the effect of the spin-orbit interaction is to lift the degeneracy between $p + ip ~(\Delta(\bveck) \sim e^{i\phi_{\bveck}})$ and $p - ip ~(\Delta(\bveck) \sim e^{-i\phi_{\bveck}})$ pairing states.
We see that in the coupled solution, each order parameter has a chirality opposite to that of its parent band ($\zeta_\lambda$), while in the decoupled one
 the chirality of $\Delta_1$ matches that of its parent band, $\zeta_1=1$.
The fact that the decoupled solution is favored suggests that the chirality of the order parameters likes to match the chirality of their respective bands.

For in-plane fields ($\delta = \pi/2$), the order parameters develop nodes. In the coupled phase both order parameters go as $\cos\phi_{\bveck}$
with nodes along the normal to the magnetic field direction, while in the decoupled phase the majority-band order parameter
goes as $\sin\phi_{\bveck}$ and has nodes along the field direction.

Before moving on, it is helpful to examine the spin structure of the pair. Because of the spin canting provided by the spin-orbit interaction, we expect
that a singlet pairing component will be mixed into each band, along with along with admixtures of spin triplet pairing \cite{Gor'kov01}.
Inverting (\ref{eq:band_tform}), we find that amplitude for creating a majority-band Cooper pair in the spin basis is
\begin{equation}
\begin{aligned}
\langle a_{\bveck1}\da a_{-\bveck1}\da\rangle = & \bigg\{ \left(1 - \frac{\alpha_R^2}{4h^2}(k_x^2 + k_y^2\cos^2\delta)\right)|\upa\upa\rangle \\ & +  \frac{\alpha_R^2}{4h^2}\left(k_x^2 - k_y^2\cos^2\delta + 2ik_xk_y\cos\delta\right)|\doa\doa\rangle \\ & + i\left(\frac{\alpha_R}{2h} - \frac{\alpha_R^3}{16h^3}\left(k_x^2+k_y^2\cos^2\delta\right)\right) \\ & \times (k_x  + ik_y\cos\delta )(|\upa\doa\rangle - |\doa\upa\rangle)\bigg\}\\ & \times (\cos\phi_{\bveck}\cos\delta + i\sin\phi_{\bveck})
\end{aligned}
\end{equation}

When the magnetic field is oriented along the SOC axis ($\delta = 0$), the dominant $|\upa\upa\rangle$ pairing shares the $p+ip$ symmetry of the majority band order parameter.
Triplet pairing of opposite spin $|\doa\doa\rangle$ is admixed with magnitude $\alpha_R^2 k^2 / h^2$, possessing a $f_{x^3 - 3xy^2} + if_{3x^2y - y^3}$ pairing symmetry.
Additionally, singlet pairing of magnitude $\alpha_R k/h + \alpha_R^3 k^3 / h^3$ and $d_{x^2-y^2}+id_{2xy}$ symmetry is realized, and so even in the decoupled phase all
types of spin pairing except triplet pairing with $S_z = 0$ exist in the system. From this, we see that the total angular momentum of the system is $j_z = 2$. We also note that as expected, increasing
the strength of SOC increases the amount of $|\doa\doa\rangle$ and $|\upa\doa\rangle - |\doa\upa\rangle$ pairing that is admixed.

When the magnetic field is completely in-plane, the dominant spin-up triplet pairing state has two nodes along the magnetic field direction, while the minority spin-down
triplet state has four nodes with $d_{x^2y}$ symmetry and the singlet state also has four nodes, obeying a $d_{xy}$ symmetry with a small admixture of $d_{x^3y}$.
Thus the triplet states share the $\varepsilon(k_x,k_y) = \varepsilon(-k_x,k_y)$ symmetry of the larger Fermi surface, while the singlet state does not.

To summarize, when the magnetic field is normal to the 2DEG plane the majority band has an isotropic gap with $p_x+i p_y$ type pairing. As the field inclines,
the $\cos\phi_{\bveck}$ component of the order parameter decreases in strength with the inclination angle $\delta$, causing the angular modulation of
the superconducting gap around the Fermi surface. Eventually, at $\delta=\pi/2$,  the order parameter acquires a
pure $p_y$ symmetry and develops nodes in the directions parallel to the magnetic field.
The minority band remains in the normal state independent of the magnetic field orientation.

\subsection{Specific heat}\label{sec:specific_heat}

\begin{figure}
\hspace{-2em}
\includegraphics[width=.45\textwidth]{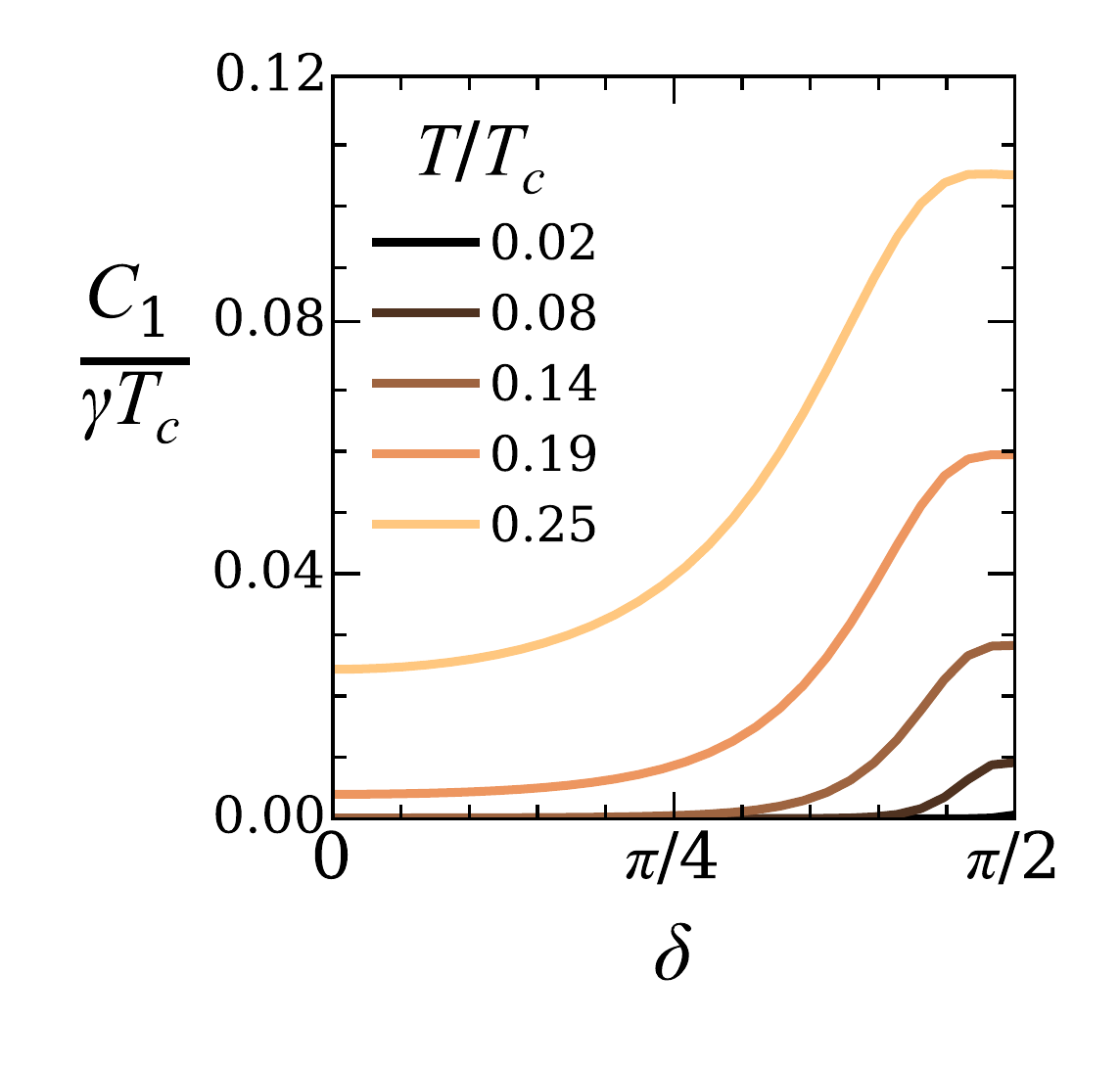}
\caption{ \label{fig:vardelta} The majority-band contribution to the specific heat $C_1$ as a function of the magnetic field inclination angle,
shown for a few different values of the ratio $T / T_c$ in the low-temperature regime. Here, $\gamma$ denotes the coefficient of the minority
band's contribution to the total specific heat, which is linear in $T$.} \end{figure}

The dependence of the pairing symmetry on the magnetic field orientation provides an ideal way for connecting our work with experiment, since the
field orientation is easily tuned in the lab simply by rotating the sample. The fact that the gap develops nodes when the magnetic field is
oriented in-plane produces signatures in a handful of experimentally accessible quantities (e.g. the $T$ dependance of the London
penetration depth\cite{Matsuda06}, the thermal conductivity\cite{Shakeripour09}, the electronic specific heat\cite{Wang01}, etc).

We focus on the electronic specific heat as an example, and calculate $C(T)$ as a function of the magnetic field orientation.
The electronic specific heat is calculated by (setting $k_B = 1$)
\begin{equation} \label{eq:specific_heat}
C(T) = \sum_{\lambda}\int_{k_{l,\lambda}}^{k_{u,\lambda}} dk~k \int_0^{2\pi} \frac{d\phi}{2\pi}\sqrt{\xi\kl^2 + 4|\Delta_{\lambda}(\bveck)|^2}\frac{\partial f_{\bveck\lambda}}{\partial T}.
\end{equation}
Here, $k_{{\rm l},\lambda}$ ($k_{{\rm u},\lambda}$) is the momentum corresponding to the energy at which we impose a lower (upper) cutoff to the energy integration.

Although the full calculation must be carried out numerically, we can estimate the behavior of $C(T)$ in the low temperature limit.
For all magnetic field orientations, the contribution from the gapless minority band is linear in $T$ as $\Delta_2=0$. The contribution to the specific heat from
the majority band depends on the type of pairing symmetry realized, and consequently on the magnetic field orientation.
When the magnetic field is parallel to the SOC axis the integral over $\phi$ can be done easily, and the total specific heat is approximated by
\begin{equation}
\label{eq:CTout}
C(T)_{\bvec{H}||\bvec{\hat z}} \sim \frac{\rho_1 \Delta_1^{5/2}}{T^{3/2}} e^{-\Delta_1/T} + \rho_2 T ,
\end{equation}
where as before, $\rho_{\lambda}$ is the density of states of the band $\lambda$.

When the magnetic field is completely in-plane the order parameter of the majority band has nodes along the field direction,
which affects the temperature dependence of the specific heat. The dominant contribution to the specific heat occurs near
the nodal points at $\phi_{\bveck} = 0,\pi$ where the gap vanishes, from which we can estimate that at low temperature,
\begin{equation}
\label{eq:CTin}
C(T)_{\bvec{H}\perp\bvec{\hat z}} \sim \frac{\rho_1 T^2}{\Delta_1}  + \rho_2 T .
\end{equation}
The dependence of the specific heat on $T^2$ is a characteristic feature of order parameters with nodal symmetry, as has
been pointed out in\cite{Movshovich01,Matsuda06}. It is worth noting here that $T^2$ behavior will also be present
for the {\em almost} in-plane orientation of the magnetic field, $\pi/2 - \delta \ll 1$, when the minimal value of the superconducting gap,
$\Delta_{\rm min} = \Delta_1 \cos\delta$ is much smaller than its maximum value $\Delta_{\rm max} = \Delta_1$. In this case \eqref{eq:CTin}
will hold in the intermediate temperature regime $\Delta_{\rm min} \ll T \ll \Delta_{\rm max}$. For $T \ll \Delta_{\rm min}$ the specific heat
will crossover to the exponentially suppressed regime of \eqref{eq:CTout}.

For general magnetic field orientations we resort to integrating \eqref{eq:specific_heat} numerically.
Figure~\ref{fig:vardelta} shows the contribution to the specific heat from the majority band, plotted against $\delta$ for
several different values of the ratio $T/T_c$ and normalized by $\gamma$, the coefficient of the (gapless) minority band's contribution to the specific heat.

The contribution from the minority band is linear in $T$, and does not affect the shape of the curves in figure~\ref{fig:vardelta}.
The difference in specific heat between the two limiting field orientations is a factor of a few, which should be easily observable in an experimental setting.

In addition to the specific heat, thermal conductivity measurements offer another way of connecting our work to experiments.
As in the case of the specific heat, the thermal conductivity will be exponentially suppressed when the magnetic field is out-of-plane and assume a
power-law dependence on $T$ when the field is in-plane. Finally, the thermal conductivity will be anisotropic when the field has an in-plane component,
making the thermal conductivity an effective probe of the pairing symmetry's dependence on the field orientation.

\section{Discussion} \label{sec:dis}

To summarize, we have investigated the role that the spin-orbit interaction plays in unconventional superconductivity  in spin-polarized two-dimensional electronic gases.

We showed that the spin-majority band has $p+ip$ pairing when the Zeeman field is out of plane of the 2DEG and parallel to the SOC axis, but develops nodes along the field direction when the field is rotated in-plane. This physics of the system can be readily probed by a measurement of the electronic specific heat.

Somewhat surprisingly, for not too large spin-orbit coupling (as compared to the Zeeman coupling), the minority band remains gapless for all magnetic field geometries, turning the considered system into an unusual non-unitary {\em half-superconductor}.

Because only the majority band is gapped, this system realizes an effectively ``spinless'' superconducting state, even though both bands have non-zero occupation. This is in contrast with other ways of engineering spinless superconductors, which usually involve carefully tuning the chemical potential so that only a single band intersects the Fermi level\cite{Alicea2012}.

Further, the system considered here support Majorana modes at its edges~\cite{SchnyderReview}. When the field has a non-zero out-of-plane component, the superconducting states possess chirality fixed by the SOC-induced lifting of the $p\pm ip$ pairing degeneracy, which is inherited by the edge modes. Purely in-plane fields lead to a nodal $p$-wave order parameter, having non-chiral flat-band edge states\cite{Nagaosa2013}. Importantly, in the present case the surface states should be thought of as surface resonances. Indeed, due to the presence of gapless states on the minority band and Rashba spin-orbit coupling, the two Fermi surfaces are in general coupled by potential disorder, of which a sample boundary is an example. Therefore, the edge states formed by the majority band are in general hybridized with the bulk states from the minority band, and thus are not sharply defined.

Our treatment is limited to the leading, $O(U^2)$, order in the electron-electron interaction strength $U$. It has been shown in Ref.~\onlinecite{Chubukov93} that the minority band becomes superconducting in the next, $O{(U^3)}$, order of the perturbative expansion. This minority band superconductivity appears due to interaction-induced corrections to the susceptibility which violate the flatness of $\chi$ for $q < 2 k_{\rm f}$, and does not require any critical threshold value of $U$. These considerations imply that, strictly speaking, our findings apply at temperatures above the critical one for the minority band, which is parametrically smaller than the critical temperature for the majority band.

A study with arbitrary SOC strength and an in-plane magnetic field was carried out in Ref.~\onlinecite{Loder13}. There and in Ref.~\onlinecite{Wang14}, it was found that inter-band
coupling can allow for a small gap on the minority band to be energetically favored for stronger SOC, suggesting that there may exist a phase transition
between the coupled and decoupled solutions at $\alpha_R k_{\rm f} \sim h$.
As such, it would be valuable to formulate SOC non-perturbatively within our model in order to better understand these phase transitions.

Despite being of a rather model nature, the considered Kohn-Luttinger-Rashba problem suggests an interesting possibility of purely electronic
mechanism of superconductivity in LAO/STO oxide interfaces. Given the reported co-existence of the superconducting and ferromagnetic orders
in this interesting system, one can imagine that a spontaneously developed ferromagnetic order promotes an exotic $p$-wave superconductivity
considered in our work. While it does appear that the prevailing view of the topic consists in ferromagnetism and superconductivity originating
from different electronic bands, and in addition the superconductivity originating from the standard phonon mechanism \cite{Michaeli2012}, the possibility
of a more exotic physics along the lines of our study should certainly be kept in mind. We would also like to note that the relative strength of
Zeeman ($\sim0.1$eV\cite{Michaeli2012}), and Rashba ($\sim 0.01$eV\cite{InterfaceSCreview}) spin splittings at magnetized LAO/STO interfaces places
them in the regime considered in this paper, and our treatment would be applicable for not too strong electron-electron repulsion, $m U\lesssim 0.1$.

Regarding perhaps the most studied $p$-wave candidate superconductor, Sr$_2$RuO$_4$ \cite{Kallin2012}, our work offers cautionary tale
as far as the question of effectiveness of the inter-band Josephson coupling is concerned. Our finding that the dominant
superconducting order parameter `inherits' chirality of the respective band appears to be quite general and is expected
to apply to a more realistic models of multi-band superconductivity in systems with strong spin-orbit interactions.

\acknowledgments
We thank A. Chubukov and R. Thomale for discussions.
This work was supported by a National Science Foundation grant No. REU-1559817 (EL and CW), and
by National Science Foundation grants No. DMR-1409089 (DAP) and No. DMR-1206774 (OAS).
\appendix*\label{sec:app}
\section{Finite momentum pairing}

For the magnetic field with an in-plane component the center-of-mass of each band is shifted by ${\bf Q} = Q \hat{y} \propto \hat{z} \times {\bf H}/|{\bf H}|$,
where $Q = m \alpha_R \sin\delta$, as we discussed below Eq.\eqref{eq:disp2} and illustrated in Figure~\ref{fig:FS}.
The shift is along the direction normal to the Zeeman field\cite{Barzykin02,Zheng13,Wu13,Qu13b,Xu14}.
It reaches maximum at the point of the topological transition when the two Fermi surfaces `touch' and when the SOC and Zeeman strengths are equal\citep{Loder13},
which lies well outside of the perturbative SOC regime studied here.

To account for finite center-of-mass-momentum (COMM) pairing, we assume the existence of two COMM
$\bvec{q}_{\lambda}$ that optimize pairing on each band\cite{Loder13}, so that pairing occurs between electrons with momenta
$\pm \bveck + \bvq_{\lambda}/2$. We see that by the $\varepsilon_{\lambda}(-k_x,k_y) = \varepsilon_{\lambda}(k_x,k_y)$ symmetry of the
(shifted) band dispersions, $\bvec{q}_{\lambda}$ cannot have an $x$-component, so that $\bvec{q}_{\lambda} = q_{\lambda}\bvec{\hat y}$.

We proceed by solving the two-particle Schrodinger equation, assuming a pair wavefunction of the form
\be |\Psi\rangle = \sum_{\lambda \bveck} \Phi^q_{\lambda}(\bveck) a\da_{\bveck+\bvq_{\lambda}/2,\lambda}a\da_{-\bveck+\bvq_{\lambda}/2,\lambda} | 0_{\lambda} \rangle, \ee
with $\Phi^q_{\lambda}(-\bveck) = - \Phi^q_{\lambda}(\bveck)$ and where $|0_{\lambda}\rangle$ denotes the ground state of the $\lambda$ band. The Schrodinger equation is
$(H_0 + H_I)|\Psi\rangle = E_q |\Psi\rangle$, where
\be H_0 =  \sum_{\lambda \bveck} \varepsilon_{\lambda}(\bveck + \bvq_{\lambda}/2) a\da_{\bveck + \bvq_{\lambda}/2,\lambda}a_{\bveck+\bvq_{\lambda}/2,\lambda} \ee
is the free Hamiltonian and $H_I$ is the interaction part. The form of the interaction matrix is calculated using the same transformation as in the $\bvq_{\lambda} = 0$ case considered earlier. Each term in $H_I$ goes as $O(U^2)$ or $O(U\alpha_R^2)$, and so any corrections to $H_I$ caused by finite COMM pairing are small enough to be dropped in our approximation scheme (also, see discussion around \eqref{eq:chifull}). Thus we have
\begin{eqnarray}
\label{eq:app1}
H_I = & \displaystyle\sum_{\lambda\mu} \displaystyle\sum_{\bveck\bveck'} g_{\lambda\mu}(\bveck,\bveck')
a\da_{\bveck+\bvq_{\lambda}/2,\lambda}a\da_{-\bveck+\bvq_{\lambda}/2,\lambda} \nonumber
\\ & \times a_{-\bveck'+\bvq_{\mu}/2,\mu}a_{\bveck'+\bvq_{\mu}/2,\mu}
\end{eqnarray}
where $g_{\lambda\mu}$ is given by (\ref{eq:intmat}). Since the interaction term is not changed from before, the decoupled solution is still energetically favored --
from here on we specialize to this case. Therefore, we can let $g_{\lambda\mu} \to \frac{U^2 \chi_{2}({\bveck}-{\bveck'})}{2V}\delta_{\lambda 1}\delta_{\lambda \mu}$.

To proceed, we take the inner product of the Schrodinger equation with
$\langle 0_{\lambda} | a_{-\bvp+\bvq_{\lambda}/2,\lambda} a_{\bvp+\bvq_{\lambda}/2,\lambda}$, obtaining
\begin{eqnarray}
&&\frac{U^2}{V}\sum_{\bveck} \chi_{2}({\bveck}-{\bvp}) \Phi^q_{1}(\bveck)  = \Phi^q_{1}(\bvp)\times\nonumber\\
&&\Big(E_q - [\varepsilon_\lambda(\bvp+\bvq_1/2) +  \varepsilon_\lambda(-\bvp+\bvq_1/2)] \Big)
\end{eqnarray}
Using \eqref{eq:disp2} the bracket on the RHS is simplified to $(E_q - [-2h - Q^2/m + (q_1 - 2Q)^2/(4m) + \bveck^2/m])\equiv - \Omega_q - 2\xi_\bveck$,
where $\Omega_q > 0$ is the two-particle bound state energy measured from the twice Fermi energy. Next, we expand both sides in angular harmonics series,
e.g. $\Phi^q_{1}(\bveck) = \sum_n f_n(k) e^{i n \phi_{\bveck}}$, to obtain
\be
-\rho U^2 \chi_2^{(-\ell)} \int_0^{\omega_c} d\xi f_\ell  = (\Omega_q + 2 \xi) f_\ell ,
\ee
which leads to
\be
-\rho U^2 \chi_2^{(-\ell)} \int_0^{\omega_c} \frac{d\xi}{\Omega_q + 2 \xi} = 1.
\ee
We obtain $\Omega_q^{(\ell)} = 2\omega_c \exp[-2/(\rho U^2 |\chi_2^{(\ell)}|)]$.
As argued previously, the highest $\Omega_q^{(\ell)}$ occurs in the $\ell=1$ channel, when  $\chi_2^{(\ell)}$ reaches the most negative value.
Therefore, the energy of the two-particle state has the form $E_q = 2 \epsilon_{\rm f} - \Omega_q^{(1)} + [(q_1 - 2Q)^2 - 4Q^2]/(4m)$
and is minimized by $q_1 = 2Q = 2 m \alpha_R \sin\delta$.

Finally, we point out that the same result can also be obtained through a mean-field analysis by making the replacement
$\xi\kl \rightarrow (\xi_{\bveck + \bvq_{\lambda}/2,\lambda} + \xi_{-\bveck + \bvq_{\lambda}/2,\lambda})/2$ in the quasiparticle dispersion $E\kl$,
and then (numerically) minimizing the thermodynamic potential $\Omega$ with respect to $\bvq_{\lambda}$.

\bibliography{current_submitted_version}

\end{document}